# Boundary integrated neural networks (BINNs) for 2D elastostatic and piezoelectric problems: Theory and MATLAB code

Peijun Zhang[1], Chuanzeng Zhang[1*], Yan Gu[2,3*], Wenzhen Qu[2,3*], Shengdong Zhao[2,3]

[1]Department of Civil Engineering, University of Siegen, Paul-Bonatz-Str. 9-11, D-57076 Siegen, Germany

[2]School of Mathematics and Statistics, Qingdao University, Qingdao 266071, PR China

[3]Institute of Mechanics for Multifunctional Materials and Structures, Qingdao University, Qingdao 266071, PR China

## Abstract

In this paper, we make the first attempt to apply the boundary integrated neural networks (BINNs) for the numerical solution of two-dimensional (2D) elastostatic and piezoelectric problems. BINNs combine artificial neural networks with the well-established boundary integral equations (BIEs) to effectively solve partial differential equations (PDEs). The BIEs are utilized to map all the unknowns onto the boundary, after which these unknowns are approximated using artificial neural networks and resolved via a training process. In contrast to traditional neural network-based methods, the current BINNs offer several distinct advantages. First, by embedding BIEs into the learning procedure, BINNs only need to discretize the boundary of the solution domain, which can lead to a faster and more stable learning process (only the boundary conditions need to be fitted during the training). Second, the differential operator with respect to the PDEs is substituted by an integral operator, which effectively eliminates the need for additional differentiation of the neural networks (high-order derivatives of neural networks may lead to instability in learning). Third, the loss function of the BINNs only contains the residuals of the BIEs, as all the boundary conditions have been inherently incorporated within the formulation. Therefore, there is no necessity for employing any weighing functions, which are commonly used in traditional methods to balance the gradients among different objective functions. Moreover, BINNs possess the ability to tackle PDEs in unbounded domains since the integral representation remains valid for both bounded and unbounded domains. Extensive numerical experiments show that BINNs are much easier to train and usually give more accurate learning solutions as compared to traditional neural network-based methods.

---

* Corresponding Authors: Yan Gu (guyan1913@163.com), Chuanzeng Zhang (c.zhang@uni-siegen.de)

## 1. Introduction

Traditional methods for solving PDEs, such as finite difference, finite element or spectral methods, often rely on grid-based discretizations and can pose computational challenges when applied to complex problems or high-dimensional systems [1-3]. In recent years, numerical methods based on deep learning techniques, such as the deep Galerkin method (DGM) [4] and physics-informed neural networks (PINNs) [5-10], have emerged as a promising alternative for solving a wide range of PDEs. The key idea behind these methods is to incorporate known physics equations (the underlying PDEs) and the corresponding initial/boundary conditions as constraints during the training of neural networks. By incorporating these physics-based constraints into the loss function, the neural network can learn to satisfy PDEs while also fitting the available data. The advantage of neural network-based methods lies in its ability to handle complex and high-order PDEs with irregular geometries and varying boundary conditions.

It's worth noting that while neural networks offer several advantages in terms of flexibility and accuracy, they also have some limitations and potential disadvantages [11-13]. Here are a few drawbacks to consider. Firstly, careful consideration should be given to the choice of network architecture, loss functions, and training strategies to ensure accurate and stable solutions. These choices are often problem-dependent and finding the optimal configuration can be challenging and time-consuming. Secondly, the training process involves solving the PDEs at every iteration to enforce the physics-informed constraints. This can result in longer training times and increased computational resources compared to traditional numerical methods. Additionally, neural network-based methods also often encounter significant challenges when tackling some intricate problems, especially those involving oscillatory or even singular solutions. This issue becomes more pronounced with higher-order derivatives, as the gradients can either become extremely small or extremely large, leading to numerical instability during the training process [14-16].

Recently, a new framework called boundary integrated neural networks (BINNs) has been introduced to effectively solve certain types of boundary value problems. BINNs combines the power of artificial neural networks with the well-established boundary element method (BEM) to effectively solve PDEs. By utilizing fundamental solutions or Green's functions, the original PDEs can be solved though boundary integral equations (BIEs) composed of a set of explicit kernels and unknown boundary quantities. A network structure is subsequently designed and trained to effectively tackle these unknown boundary quantities, where the output of the neural networks can satisfy the original PDEs automatically. As a result, the learning process is remarkably accelerated and fortified, ensuring enhanced efficiency and reliability.

In contrast to traditional neural network-based methods, BINNs offer several distinct advantages. First, BINNs seamlessly integrate prior knowledge about BIEs into the training process, allowing for a focus solely on fitting the boundary conditions. This significantly reduces the training data required, resulting in a faster and more efficient learning process. Second, the substitution of the differential operator in the PDEs with an integral operator effectively eliminates the requirement for further differentiation of the neural networks. This substitution is particularly beneficial as high-order derivatives of neural networks can introduce instability during the learning process. At last, BINNs possess the ability to tackle PDEs in unbounded domains since the integral representation remains valid for both bounded and unbounded domains.

Recently, Lin et al. [14] made a pioneering contribution by employing neural networks to approximate the solutions of an indirect BIEs characterized by pure Dirichlet boundary conditions. In their work, they devised and trained the network structure based on the principles of single- and double-layer potential theories. Another related research was conducted by Zhang et al. [17] where they investigated the feasibility of employing artificial neural networks to solve direct BIEs with mixed boundary conditions. Additionally, they utilized isogeometric BEM techniques and employed Non-Uniform Rational B-Splines (NURBS) to parameterize the boundary. By incorporating these tools, they were able to effectively model and represent the complex geometry of the boundary in their neural network-based approach. The study described in [17] focused solely on the Laplacian equation, without extending their analysis to encompass more general classes of equations. In a more recent study, Sun et al. [18] documents the first attempt to apply the BINNs for the numerical solutions of elastostatic problems.

In this study, we present the initial application of BINNs for solving two-dimensional (2D) elastostatic and piezoelectric problems, which involve the study of stresses and deformations in elastic materials and the interaction between mechanical and electrical fields in certain materials, respectively. By making the first attempt to apply BINNs to these problems, our objective is to investigate the viability and efficacy of this approach in resolving problems across diverse scientific and engineering contexts. The rest of the paper is organized as follows. Section 2 provides an overview of the application of boundary integral method for solving general elastostatic and piezoelectric problems. The artificial neural networks and the computational methodology of the present BINNs are describes in Section 3. This section also outlines the data gathering and analysis procedures undertaken to ensure the validity and reliability of the findings. In Section 4, we present extensive numerical experiments, and the obtained results are compared to existing analytical or reference solutions obtained using traditional neural network-based methods. Section 5 concludes the paper by summarizing the main findings and providing additional remarks. Additionally, for further reference, a self-contained MATLAB code is included in the Supplementary material.

## 2. Problem Statement and Boundary Integral Equations (BIEs)

### 2.1. Problem Statement

The governing equations of linear elasticity, without the body force, can be expressed as:

$$\sigma_{ij,j} = 0, \quad i = 1, 2, \tag{1}$$

with the boundary conditions:

$$u_i = \bar{u}_i, \quad \text{on } \Gamma_u, \tag{2}$$

$$t_i = \sigma_{ij} n_j = \bar{t}_i, \quad \text{on } \Gamma_t, \tag{3}$$

where $\sigma_{ij}$, $u_i$ and $t_i$ represent the stress, displacement and boundary traction components, respectively, $n_j$ denotes the unit outward normal vector, and $\Gamma_u$ and $\Gamma_t$ refer to the boundary sections where the displacement and traction conditions are specified. Here and in what follows, the customary Euler notation for summation over repeated subscripts is utilized.

Based on the theory of linear piezoelectricity, the coupled mechanical and electrical equilibrium equations for piezoelectric materials, in the absence of body force, can be expressed as follows [19-21]:

$$\sigma_{ij,j} = 0, \quad i, j = 1, 2, \tag{4}$$

$$D_{i,i} = 0, \quad i = 1, 2, \tag{5}$$

with the boundary conditions:

$$\begin{cases} u_i = \bar{u}_i & \text{on } \Gamma_u \\ \sigma_{ij}n_j = \bar{t}_i & \text{on } \Gamma_t \end{cases} \quad \text{and} \quad \begin{cases} \phi = \bar{\phi} & \text{on } \Gamma_\phi \\ D_i n_i = -\bar{\omega} & \text{on } \Gamma_\omega \end{cases}, \tag{6}$$

where again $\sigma_{ij}$, $u_i$ and $t_i$ represent the stress, displacement and traction components, respectively, $D_i$ stands for the electric displacement vector, $\omega$ and $\phi$ denote the surface charge and the electric scalar potential, respectively. The aforementioned equations constitute the comprehensive mathematical representation of the elastostatic and piezoelectric fields in 2D solids. For further details, interested readers are referred to Refs. [19-21].

## 2.2. Boundary Integral Equations (BIEs)

By applying Green's identities, the initial PDEs can be transformed into a set of BIEs. The universal representation of the BIEs, applicable to both elastostatic and piezoelectric equations, can be expressed as follows:

$$\mathbf{C}(P)\mathbf{u}(P) = \int_S \mathbf{U}(P,Q)\mathbf{t}(Q)dS(Q) - \fint_S \mathbf{T}(P,Q)\mathbf{u}(Q)dS(Q), \tag{7}$$

with the notations:

$$\mathbf{u} = \begin{bmatrix} u_1 \\ u_2 \end{bmatrix}, \quad \mathbf{t} = \begin{bmatrix} t_1 \\ t_2 \end{bmatrix}, \quad \mathbf{U} = \begin{bmatrix} U_{11} & U_{12} \\ U_{21} & U_{22} \end{bmatrix}, \quad \mathbf{T} = \begin{bmatrix} T_{11} & T_{12} \\ T_{21} & T_{22} \end{bmatrix}, \tag{8}$$

for linear elastic body [22, 23] and

$$\mathbf{u} = \begin{bmatrix} u_1 \\ u_2 \\ -\phi \end{bmatrix}, \quad \mathbf{t} = \begin{bmatrix} t_1 \\ t_2 \\ -\omega \end{bmatrix}, \quad \mathbf{U} = \begin{bmatrix} U_{11} & U_{12} & U_{13} \\ U_{21} & U_{22} & U_{23} \\ U_{31} & U_{32} & U_{33} \end{bmatrix}, \quad \mathbf{T} = \begin{bmatrix} T_{11} & T_{12} & T_{13} \\ T_{21} & T_{22} & T_{23} \\ T_{31} & T_{32} & T_{33} \end{bmatrix}, \tag{9}$$

for piezoelectric solid [19, 20, 24].

In the above equations, $P$ and $Q \in \partial\Omega$ are observation and source points, respectively, $\{u_i\}_{i=1,2}$ and $\{t_i\}_{i=1,2}$ denote the mechanical displacement and traction components, respectively. The jump term $\mathbf{C}(P)$ solely relies on the local geometry at point $P$ with a specific value of $C_{ij}(P) = \delta_{ij}/2$ assigned to a smooth boundary, $\fint_S$ represents the singular integral with strong-singularities in the context of Cauchy principal value (CPV), **U** and **T** are fundamental solutions (Green's functions). For 2D elastostatic problems, the fundamental solutions are given by:

$$U_{ij}(P,Q) = \frac{1}{8\pi\mu(1-v)}\left[(4v-3)\ln(r)\delta_{ij} + r_{,i}r_{,j}\right], \tag{10}$$

$$T_{ij}(P,Q) = -\frac{1}{4\pi(1-v)r}\left\{\frac{\partial r}{\partial \boldsymbol{n}}\left[(1-2v)\delta_{ij} + 2r_{,i}r_{,j}\right] - (1-2v)(r_{,i}n_j - r_{,j}n_i)\right\}, \tag{11}$$

where $\mu$ represents the shear modulus while $v$ corresponds to Poisson's ratio, $\boldsymbol{n} = (n_1, n_2)$ represents the unit outward normal vector at point $Q$, $r = |P\text{-}Q|$ stands for the Euclidean distance between points $P$ and $Q$. The explicit formulations of fundamental solutions for piezoelectric equations can be found in Refs. [19, 20, 24].

It is worth mentioning that kernel function with $\mathbf{U}(P,Q)$ exhibits a weak-singularity of order $O(\ln r)$, whereas the kernel with $\mathbf{T}(P,Q)$ displays a strong-singularity of order $O(r^{-1})$. To evaluate integrals involving various orders of singularities, numerous direct and indirect algorithms have been developed within the BEM community. These methods can also be applied within the framework of the BINNs approach proposed in this study. For instance, to compute integrals with a strong singularity, we employ a singularity subtraction technique proposed by Guiggiani and Casalini [25]. This technique has been adopted in our calculations for its effectiveness and reliability. A comprehensive discussion on the computation of singular integrals goes beyond the scope of this paper. For further details, interested readers are referred to Refs. [26-29].

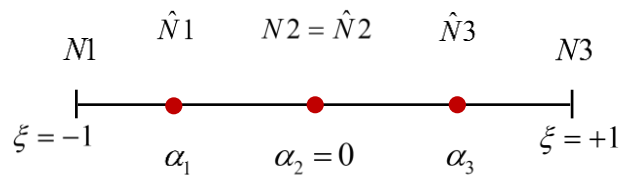

**Fig. 1.** Local coordinate system for discontinuous quadratic elements.

In the BEM framework, the boundary $\partial\Omega$ of the domain is firstly divided into a collection of $N$ segments, not necessarily equal, which are known as the boundary elements. Two approximations are utilized for each segment, one regarding the boundary's geometry and the other concerning the variations in physical quantities. Without loss of generality, the geometry of the boundary is approximated here using traditional quadratic elements as follows:

$$x_i = x_i^1 N_{1G}(\xi) + x_i^2 N_{2G}(\xi) + x_i^3 N_{3G}(\xi), \quad i=1,2, \tag{12}$$

where $\{x_i\}_{i=1,2}$ represents the $i^{\text{th}}$ coordinate component of an arbitrary boundary point, $\{x_i^k\}_{k=1}^{3}$ denote coordinate components of the interpolation points $N1$, $N2$ and $N3$ (see Fig. 1), and

$$N_{1G}(\xi) = \xi(\xi-1)/2, \quad N_{2G}(\xi) = 1-\xi^2, \quad N_{3G}(\xi) = \xi(\xi+1)/2, \tag{13}$$

stand for quadratic interpolation functions. In order to avoid the well-known corner problem, the physical quantities $\mathbf{u}$ and $\mathbf{t}$ along the boundary are approximated by using the following discontinuous quadratic elements [22]:

$$\mathbf{u} = \mathbf{u}^1 N_{1P}(\xi) + \mathbf{u}^2 N_{2P}(\xi) + \mathbf{u}^3 N_{3P}(\xi), \tag{14}$$

$$\mathbf{t} = \mathbf{t}^1 N_{1P}(\xi) + \mathbf{t}^2 N_{2P}(\xi) + \mathbf{t}^3 N_{3P}(\xi), \tag{15}$$

where $\{\mathbf{u}^i\}_{i=1}^{3}$ and $\{\mathbf{t}^i\}_{i=1}^{3}$ are physical quantities at interpolation points $\hat{N}1$, $\hat{N}2$ and $\hat{N}3$, respectively (see Fig. 1), and

$$N_{1P}(\xi) = \frac{\xi(\xi-\alpha_3)}{\alpha_1(\alpha_1-\alpha_3)}, \quad N_{2P}(\xi) = \frac{(\xi-\alpha_1)(\xi-\alpha_3)}{\alpha_1\alpha_3}, \quad N_{3P}(\xi) = \frac{\xi(\xi-\alpha_1)}{\alpha_3(\alpha_3-\alpha_1)}, \tag{16}$$

represent the set of discontinuous interpolation functions in which $\alpha_1 \in (-1,0)$ and $\alpha_3 \in (0,1)$ correspond to the local coordinates of the interpolation points $\hat{N}1$ and $\hat{N}3$, respectively. In our computations, we have utilized the values $\alpha_1 = -0.8$ and $\alpha_1 = 0.8$, respectively, but other choices for these values are also possible.

Based on the aforementioned notations, we can now discretize the boundary integral equation

(7) in the following manner:

$$\mathbf{C}(P^m)\mathbf{u}(P^m) = \sum_{n=1}^{N}\sum_{k=1}^{3}\mathbf{t}^{n,k}\int_{S_n}\mathbf{U}\left(P^m,Q_n(\xi)\right)N_{kP}(\xi)J(\xi)d\xi \\ -\sum_{n=1}^{N}\sum_{k=1}^{3}\mathbf{u}^{n,k}\oint_{S_n}\mathbf{T}\left(P^m,Q_n(\xi)\right)N_{kP}(\xi)J(\xi)d\xi, \quad (17)$$

where $N$ represents the total number of boundary elements, $P^m$ stands for the $m^{\text{th}}$ collocation/observation point along the boundary, $\mathbf{t}^{n,k}$ and $\mathbf{u}^{n,k}$ are physical quantities at the $k^{\text{th}}$ interpolation point of the $n^{\text{th}}$ boundary element, $J(\xi)$ is the Jacobian of the transformation from global coordinates to local coordinates. At a boundary point, only one of the boundary quantities, either $\mathbf{u}$ or $\mathbf{t}$, is prescribed based on known boundary conditions. By solving the BIEs (17), one can obtain all the unknown physical quantities along the boundary. After that, the solutions of the original PDEs at any point $P$ inside the domain $\Omega$ (not on the boundary) can be calculated as:

$$\mathbf{u}(P) = \int_{S}\mathbf{U}(P,Q)\mathbf{t}(Q)dS(Q) - \int_{S}\mathbf{T}(P,Q)\mathbf{u}(Q)dS(Q). \quad (18)$$

The above is the basic principle of the traditional BEM for solving a given boundary value problems. While various BEM-based frameworks enjoy the advantages of easy-meshing and high accuracy, their efficiency in solving large-scale problems has been a significant challenge. This is primarily because the traditional BEM formulation generates dense and non-symmetric matrices. Although these matrices are smaller in size, they still require a significant amount of memory and additional computational operations for computation when using direct solvers. Although this bottleneck can be overcome through the development of various acceleration techniques [30-32], it often involves complex analytical manipulations that are not convenient for researchers to utilize.

## 3. Artificial Neural Networks and Deep Learning for BIEs

In this section, we introduce the basic principle of the proposed boundary integrated neural networks (BINNs). Instead of solving the integral equation (17) directly, we firstly construct a neural network to approximate the unknown boundary quantities $\mathbf{u}(Q)$ and $\mathbf{t}(Q)$ through the inputs $x_1$ and $x_2$. The network is then trained to satisfy both the given boundary conditions and the integral equations (17) simultaneously at a set of collocation/observation points. To provide a clear illustration of the approach, we here focus solely on 2D elastostatic problems. Similar methodologies can be applied for addressing a variety of other problems.

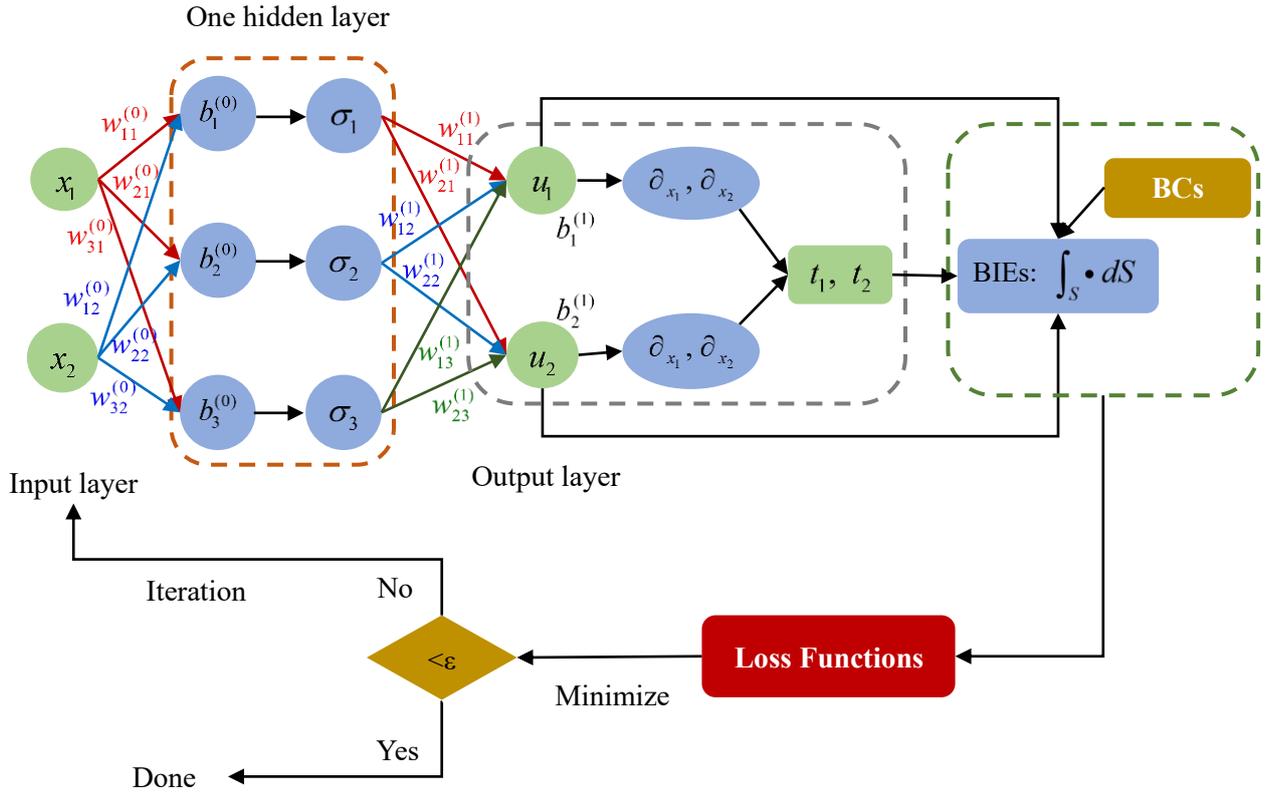

**Fig. 2.** The framework of BINNs for 2D elastostatic problems.

In the framework of artificial Neural Networks, the simplest network architecture consists of three operations: (1) the first operation with two channels corresponding to the inputs $x_1$ and $x_2$; (2) a fully-connected hidden layer with $M$ neuron units; and (3) the last operation with two output channels for the target solutions $u_1(x_1, x_2)$ and $u_2(x_1, x_2)$, where for 2D elastostatic problems $u_1$ and $u_2$ correspond to displacement components along the $x$- and $y$-axes, respectively.

The framework of BINNs for solving 2D elastostatic problems is shown in Fig. 2. In this manner,

the target solutions (displacement components) $u_1$ and $u_2$ can be approximated using neural networks as follows:

$$u_1(x_1, x_2, \boldsymbol{w}, \boldsymbol{b}) = \sum_{i=1}^{M} w_{1i}^{(1)} \sigma\left(z_i^{(0)}\right) + b_1^{(1)}, \tag{19}$$

$$u_2(x_1, x_2, \boldsymbol{w}, \boldsymbol{b}) = \sum_{i=1}^{M} w_{2i}^{(1)} \sigma\left(z_i^{(0)}\right) + b_2^{(1)}, \tag{20}$$

with

$$z_i^{(0)} = w_{i1}^{(0)} x_1 + w_{i2}^{(0)} x_2 + b_i^{(0)}, \tag{21}$$

where $w_{ij}^{(0)}$ represents the weight linking the input unit $x_j$ to the $i^{\text{th}}$ neuron unit, $w_{ki}^{(1)}$ denotes the weight from the $i^{\text{th}}$ neuron unit to the target solution $u_k$, $\{b_i^{(0)}\}_{i=1}^{M}$ and $\{b_k^{(1)}\}_{k=1}^{2}$ stand for biases associated with the $M$ neuron units and two target solutions, respectively, and $\sigma(\cdot)$ is the activation function. Since there is no regularity requirement on target solutions, various well-known and widely used activation functions, such as Sigmoid, Tanh, Swish, Softplus, Arctan, Mish and ReLU, can be utilized. The extension of the neural network approximation to cases involving multiple hidden layers can be derived in a similar manner [9, 10, 33, 34]. To approximate the unknown fluxes $\mathbf{t}(Q)$, it is necessary to calculate the first-order derivatives of the target solutions (19) and (20) with respect to the network inputs $x_1$ and $x_2$. In our computations, this is achieved by utilizing the automatic differentiation function "dlgradient" provided by the Deep Learning Toolbox in MATLAB.

**Remark 1.** In the case of mixed boundary conditions, a single/same neural network is used here across the entire boundary, where the unknown fluxes are calculated through the utilization of automatic differentiation techniques. It is worth highlighting that only the calculation of first-order derivatives of target solutions is required. This is one of significant advantages of the method since computing high-order derivatives of the neural networks can introduce instability during the learning process. However, for problems involving discontinuity, it is recommended to use separate neural networks for each part of the boundary. This strategy has been demonstrated to substantially improve the overall accuracy of the solutions [17].

In the BINNs, the trainable parameters $w$ and $b$ can be trained by substituting the target solutions $u_1$, $u_2$ and/or $t_1$, $t_2$ into the integral equations (17) and minimizing the following loss function:

$$L(w,b) = L_{\text{BIE}}^{(1)}(w,b) + L_{\text{BIE}}^{(2)}(w,b), \tag{22}$$

with the notations:

$$L_{\text{BIE}}^{(1)}(w,b) = \frac{1}{N_{total}} \sum_{m=1}^{N_{total}} \left| \frac{u_1(P^m)}{2} - \int_S U_{1j} t_j dS + \oint_S T_{1j} u_j dS \right|^2, \tag{23}$$

$$L_{\text{BIE}}^{(2)}(w,b) = \frac{1}{N_{total}} \sum_{m=1}^{N_{total}} \left| \frac{u_2(P^m)}{2} - \int_S U_{2j} t_j dS + \oint_S T_{2j} u_j dS \right|^2, \tag{24}$$

where $L_{\text{BIE}}^{(1)}(w,b)$ and $L_{\text{BIE}}^{(2)}(w,b)$ represent two loss terms corresponding to the BIEs formulated for solving 2D elastostatic problems, and $N_{total}$ stands for the total number of collocation/ observation points located along the boundary.

For problems related to 2D and 3D solid mechanics, the physical units of the loss terms could be different, leading to the residuals from the underlying PDEs and boundary conditions being at different scales or orders of magnitude [11]. During the training process, these terms may compete with each other, which can result in biased training. Although various adaptive learning schemes have been proposed to balance different loss terms automatically [7, 34-36], how to balance the scales between different loss terms still remains a challenging and open question.

**Remark 2.** In the current BINNs, the specified boundary conditions have been automatically incorporated into the BIEs, i.e., only the boundary unknowns are approximated and the network itself does not have to satisfy the given boundary conditions. Therefore, there is no necessity for employing any weighs into the loss function, which are commonly used in traditional methods to balance the gradients among different objective functions (including the terms involving the underlying PDEs and other terms corresponding to the boundary conditions). This represents another significant advantage of the present BINNs approach.

Now, the initial BIEs outlined in equation (17) are recast into an optimization problem, aiming to find a set of parameters $\boldsymbol{w}$ and $\boldsymbol{b}$ such that the loss function (22) is as small as possible. This objective can be achieved by employing various gradient-based optimization techniques. Once the trainable parameters are calculated, the unknown quantities along the boundary can be approximated by utilizing the neural networks (19) and (20).

**Remark 3.** By embedding BIEs into the learning procedure, BINNs only need to discretize the boundary of the solution domain and train the neural networks with a small number of training data, which can lead to faster and more stable learning process. Similar conclusions can be drawn from the subsequent numerical examples.

**Remark 4.** In the BEM-based methods, the final solutions are obtained by solving a system of linear equations. However, the coefficient matrices that arise in the BEM are usually dense and asymmetric, which presents a significant challenge in efficiently solving large-scale problems. Conversely, the present BINNs circumvent this issue by resolving solutions through a comprehensive training process, utilizing a variety of gradient descent-based algorithms.

## 4. Numerical Examples and Discussions

In this Section, we utilize the current BINNs to compute a range of numerical examples. The first two examples are relevant to elastostatic problems, while the third and fourth examples are specifically designed for piezoelectric problems. The influence of the network architecture, including the number of hidden layers, the number of neuron units used in each layer, and the total number of boundary elements, on the overall accuracy of the proposed BINNs is thoroughly investigated. The results of the BINNs are also compared with those obtained by using the BEM and traditional neural network-based methods. In all benchmark examples studied here, the total number of input data used to train the networks is relatively small, ranging up to a few thousand boundary points. Unless otherwise specified, all numerical integrations are performed using a 16-point Gaussian quadrature formula. All numerical experiments are tested using MATLAB R2023a and on a 64-bit Windows system with an i7 2.90 GHz CPU and 32 GB of memory. For error analysis, the following definition of relative error is employed:

$$Error = \left| \frac{u_{num} - u_{exact}}{\max u_{exact}} \right|, \qquad (25)$$

where $u_{exact}$ represents the exact/reference solution, and $u_{num}$ is the corresponding neural network approximation.

### *4.1. An elastic beam under pure bending*

Firstly, as shown in Fig. 3, we examine a simple case of a 2D plane-strain pure bending beam problem. This case study has also been studied in Ref. [11] by using the PINN-based model. The length and height of the beam are taken to be $L = 1\,\text{m}$ and $H = 0.1\,\text{m}$, respectively. A bending moment $M = \frac{1}{12}\,\text{N}\cdot\text{m}$ is applied on the side of the beam. For numerical simulation, as shown in Fig. 3, a linearly varying force $\bar{t} = 1000 \cdot y$ is subjected on the right-hand side of the beam.

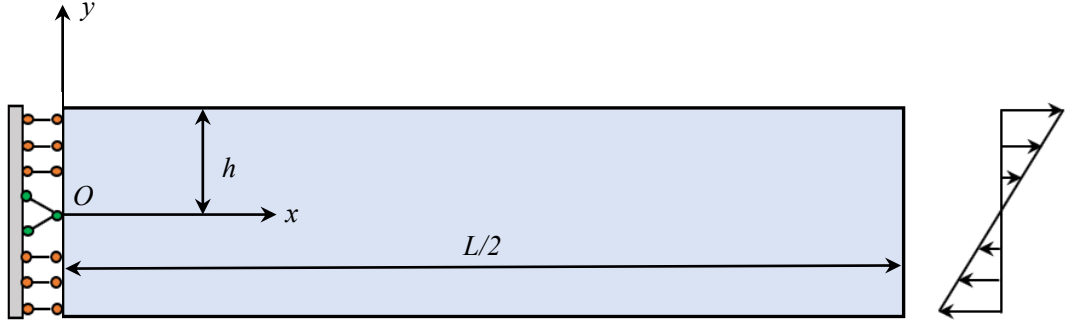

**Fig. 3.** Configuration of half pure bending beam.

The displacement boundary conditions are given as follows:

$$u_1(0, y) = 0, \quad u_2(0, 0) = 0. \tag{26}$$

Training the networks of BINNs involves utilizing a series of collocation points along the given boundaries to fulfill both the boundary conditions and the underlying BIEs. Here, we use a total of 180 discontinuous quadratic elements with 540 nodes as collocation points to facilitate the training of BINN model. To minimize the loss function, we apply 3 fully connected hidden layers, each consisting of 20 neurons. The training process runs for 5000 iterations. For each hidden layer, the hyperbolic tangent activation function "*tanh*" is applied. Plots of the calculated mechanical displacements $u_1$ and $u_2$ at arbitrary selected points of the entire boundaries are illustrated in Fig. 4. For comparison purposes, the corresponding analytical solutions given in Ref. [11] are also included. It is noteworthy that the results obtained with the present BINNs show excellent agreement with the analytical solutions. Fig. 5 illustrates the contour plots of the calculated displacements and stresses at points across the entire computational domain. The corresponding relative errors compared to the analytical solutions are also provided. It is evident that the predicted values from the current BINN demonstrate remarkable agreement with the exact solutions, with the largest error being less than 4.72E-05.

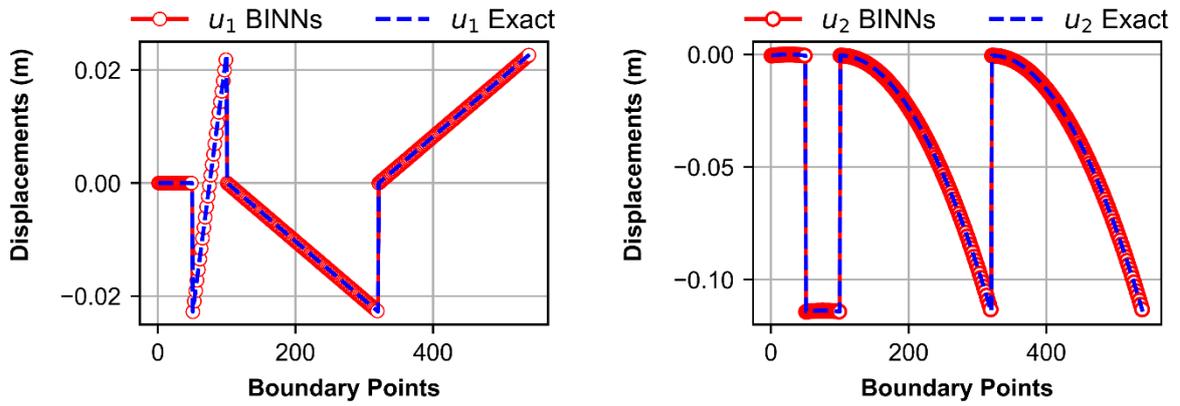

**Fig. 4.** Comparing Results: BINN and Analytical Solutions on boundaries.

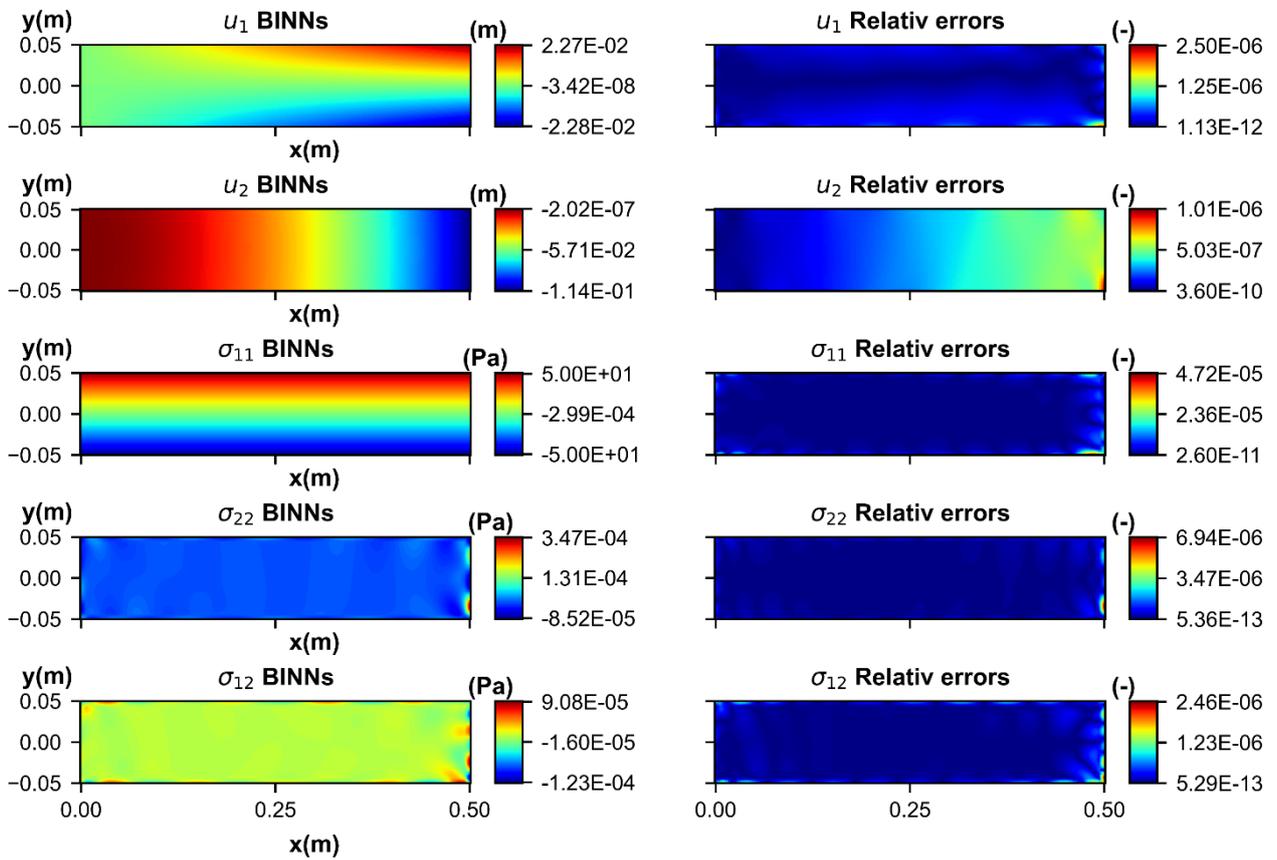

**Fig. 5.** Displacement and stress fields of the pure bending problem.

## 4.2. An elastic plane with holes under tensile loading

As shown in Fig. 6, the second example considers a $1.0\,\text{m} \times 1.0\,\text{m}$ elastic plane with 3 randomly distributed circular holes of radius $R = 0.2\,\text{m}$, with the coordinates of their centers being (-0.5, 0.5), (-0.25, -0.5), and (0.5, 0), respectively. The Young's modulus and Poisson's ratio of the material are taken as $E = 2 \cdot 10^3 \ \text{N} \cdot \text{m}^{-2}$ and $v = 0.3$, respectively. The boundary conditions are specified are as follows:

$$t_1 = 0, \quad t_2 = 0, \quad (\text{left and right boundaries}), \tag{27}$$

$$u_1(x=0) = 0, \quad t_1(x \neq 0) = 0, \quad u_2 = 0, \quad (\text{lower boundary}), \tag{28}$$

$$t_1 = 0, \quad t_2 = 1.0, \quad (\text{upper boundary}), \tag{29}$$

and, on the boundaries of all circular holes, the boundary conditions are traction-free.

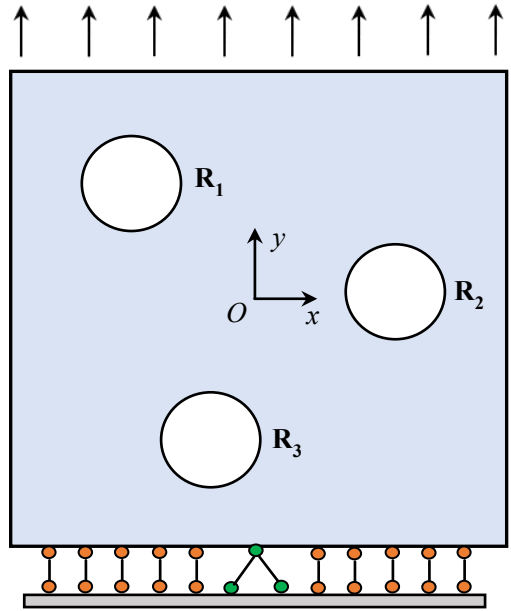

**Fig. 6.** Configuration of an elastic plane with three holes.

Here, we discretized each outer boundary with 20 discontinuous quadratic elements and each inner circular boundary with 15 elements, giving a total of 375 nodes as collocation points to facilitate BINN model training. To minimize the loss function, we used 4 fully connected hidden layers, each consisting of 30 neurons. The activation function used for each hidden layer was the Swish function, denoted as $\sigma(z) = z/(1+e^{-z})$. Fig. 7 illustrates a comparison of the calculated displacements at arbitrary selected points on three inner circular boundaries between BINNs and FEM software

ANSYS. It can be observed that the results obtained from the present BINNs agree remarkably well with the corresponding FEM solutions.

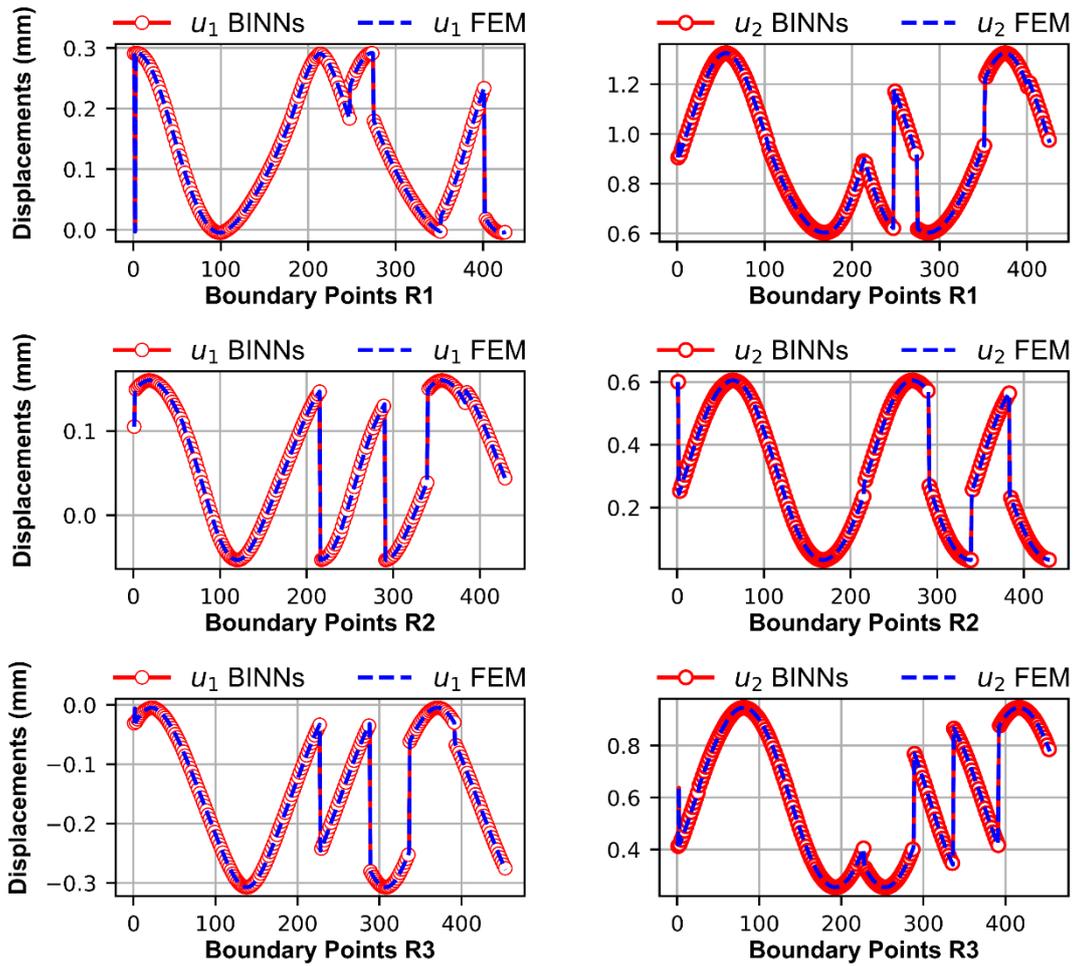

**Fig. 7.** Comparing Results: BINN and FEM Solutions on circular boundaries.

Fig. 8 illustrates the contour plots of the calculated displacements and stresses at points across the entire computational domain. Here we consider the numerical solution from the FEM as the "reference/exact solution" and compares it with the algorithm proposed in this paper. Remarkably, the displacements predicted by the present BINN demonstrate exceptional agreement with the FEM solutions, with the largest error limited to 3.24E-5. Similarly, for the stresses, the error remains confined to a mere 5E-3. Tt is reported that, as expected, refining the mesh in ANSYS could lead to further reductions in the error between BINN and FEM solutions.

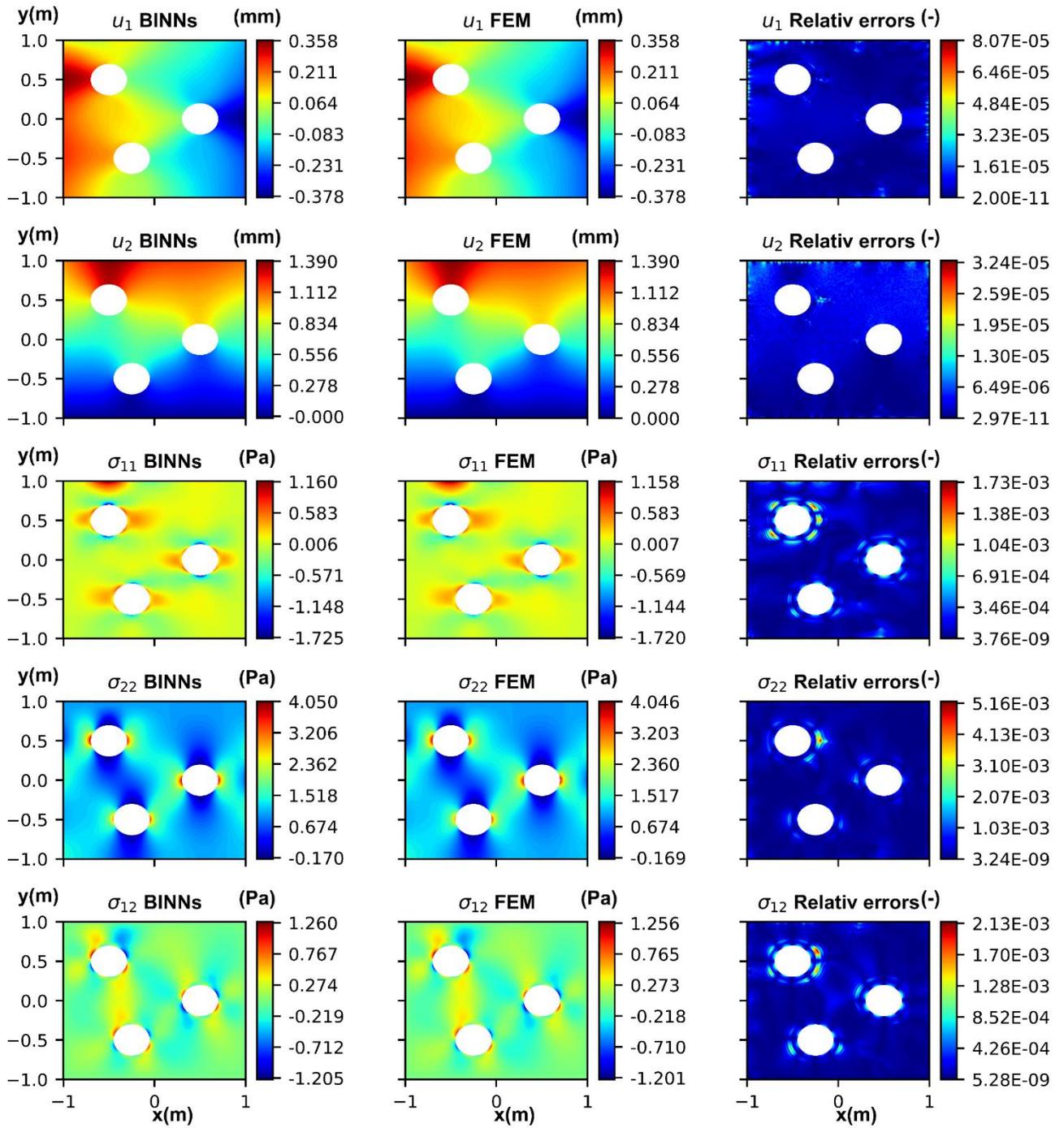

**Fig. 8.** Comparing Results: BINN and FEM Solutions for example 2.

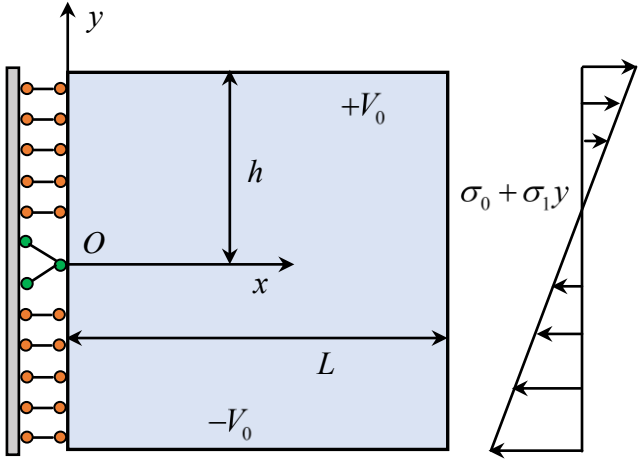
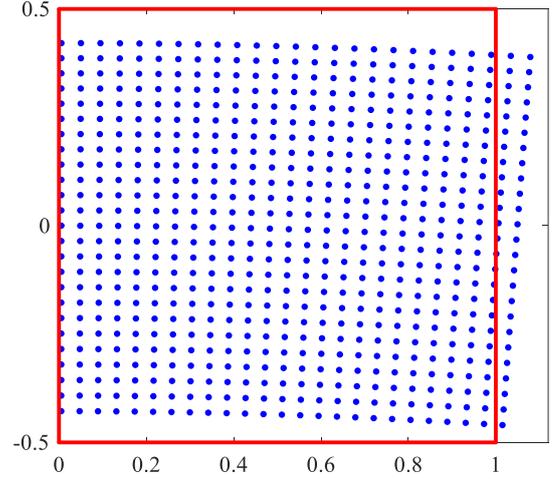

**Fig. 9.** Piezoelectric strip under linearly varying stress.

**Fig. 14.** Mechanical deformation of the piezoelectric strip (displacements are amplified by a factor of 200).

**Table 1.** The mechanical and electric constants of the considered PZT-4 ceramics [37].

| | |
|---|---|
| Elastic compliance constants | $s_{11} = 16.4$, $s_{13} = -7.22$, $s_{33} = 18.8$, $s_{55} = 47.5$ ($\times 10^{-6}$ mm$^2 \cdot$N$^{-1}$) |
| Piezoelectric constants | $d_{31} = -172$, $d_{33} = 374$, $d_{15} = 584$ ($\times 10^{-9}$ mm$\cdot$V$^{-1}$) |
| Dielectric constants | $\zeta_{11} = 1.53105$, $\zeta_{33} = 15.05$ ($\times 10^{-8}$ N$\cdot$V$^{-2}$) |

## 4.3. A piezoelectric strip under linearly varying stress and applied voltage

In the third example, as depicted in Fig. 9, we investigate the bending deformation of a 1.0×1.0mm piezoelectric strip (PZT-4) [37, 38], where the strip is subjected to a linearly varying stress in the *x*-direction, while a voltage is applied to the top and bottom surfaces of the strip. The mechanical and electric constants [38] of the considered PZT-4 ceramics can be found in Table 1. The following boundary conditions are specified along the strip surfaces:

$$u_1(0, y) = 0, \quad t_2(0, y \neq 0) = 0, \quad \phi_{,1}(0, y) = 0, \quad u_2(0,0) = 0, \tag{30}$$

$$t_1(1, y) = \sigma_0 + \sigma_1 y, \quad t_2(1, y) = 0, \quad \phi_{,1}(1, y) = 0, \tag{31}$$

$$t_1(x, \pm 0.5) = 0, \quad t_2(x, \pm 0.5) = 0, \quad \phi(x, \pm 0.5) = \pm V_0, \tag{32}$$

where $\sigma_0 = -5.0 \text{ N} \cdot \text{mm}^{-2}$, $\sigma_1 = 20.0 \text{ N} \cdot \text{mm}^{-2}$ and $V_0 = 1000 \text{ V}$. The corresponding analytical solutions for this numerical example can be find in Ref. [37].

For multi-physics problems, the physical units of the material constants usually are at different scales or in different orders of magnitude. Consequently, this can lead to a competitive interplay between the loss terms during the training, potentially may introduce biases in the learning process. To alleviate the scale-difference problem, the units of the material constants are adjusted here to a comparable magnitude by using the following scaling factors:

$$\tilde{N} = 10^6 \text{N}, \quad \tilde{V} = 10^6 \text{V}, \tag{33}$$

and the new material constants are taken to be:

$$s_{11} = 16.4, \quad s_{13} = -7.22, \quad s_{33} = 18.8, \quad s_{55} = 47.5 \quad (\text{mm}^2 \cdot \tilde{N}^{-1}), \tag{34}$$

$$d_{31} = -0.172, \quad d_{33} = 0.374, \quad d_{15} = 0.584 \quad (\text{mm} \cdot \tilde{V}^{-1}), \tag{35}$$

$$\zeta_{11} = 0.0153105, \quad \zeta_{33} = 0.1505 \quad (\tilde{N} \cdot \tilde{V}^{-2}). \tag{36}$$

Training the BINNs requires a set of collocation points located along the boundary in order to ensure the satisfaction of the BIEs and the corresponding boundary conditions. Here, we utilize a total number of $N = 60$ discontinuous quadratic elements, resulting in 180 collocation points, to train the present BINNs model. We minimize the loss function here through the utilization of 2 fully-connected hidden layers with 10 neurons per layer and 2500 iteration steps. For each hidden layer, the activation function employed is the Swish function, denoted as $\sigma(z) = z / (1 + e^{-z})$.

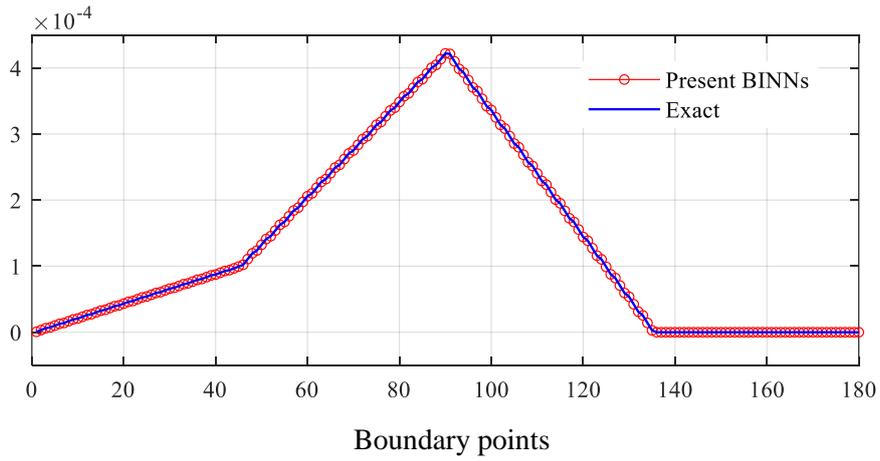

**Fig. 10.** Results of displacements $u_1\,(\mathrm{mm})$ at points along the boundary.

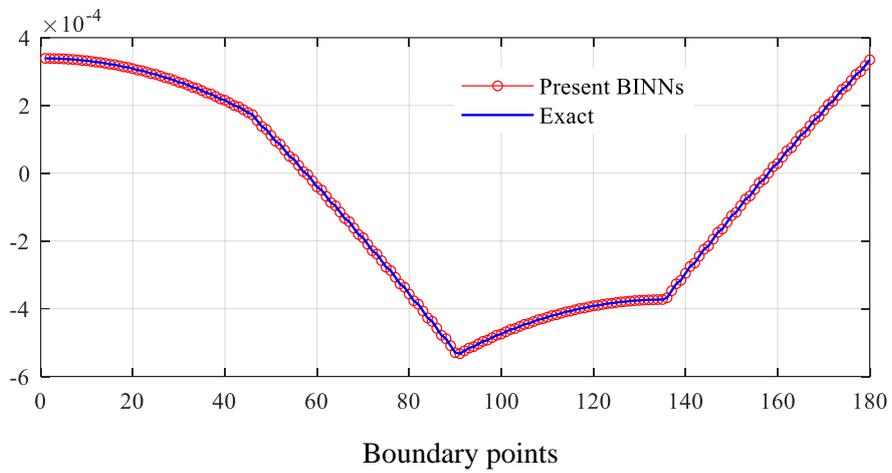

**Fig. 11.** Results of displacements $u_2\,(\mathrm{mm})$ at points along the boundary.

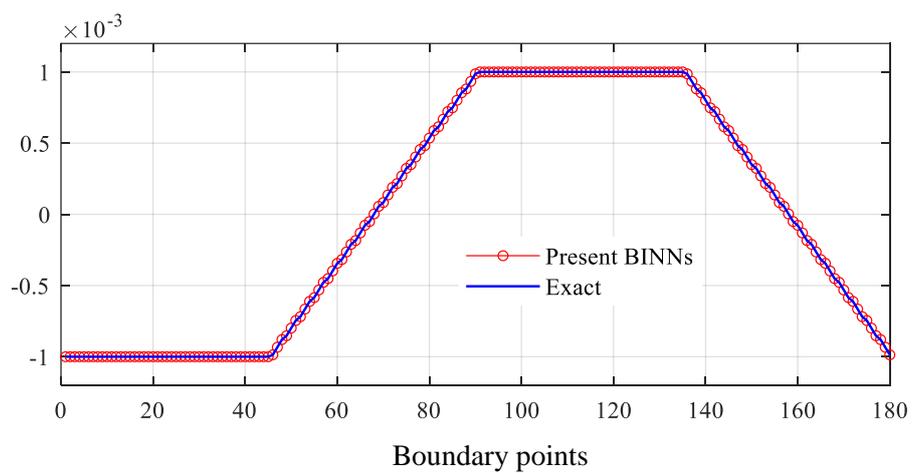

**Fig. 12.** Results of electric potential $\phi\,(\tilde{\mathrm{V}})$ at points along the boundary.

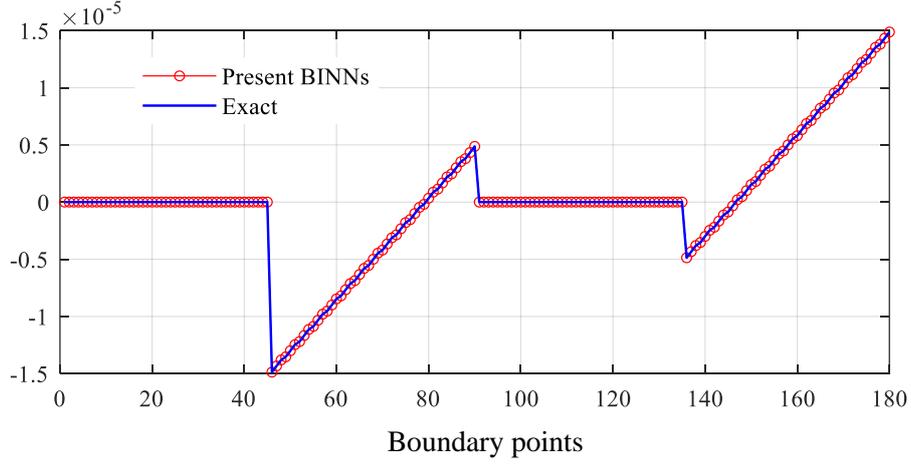

**Fig. 13.** Results of boundary traction $t_1\left(\tilde{\mathrm{N}}\cdot\mathrm{mm}^{-2}\right)$ at points along the boundary.

Plots of the computed mechanical displacements $u_1$, $u_2$ and electric potential $\phi$ at points along the entire boundary are given in Figs. 10-12, respectively, where the exact solutions are also given for the purpose of comparison. It can be observed that the results obtained from the present BINNs agree remarkably well with the corresponding analytical solutions. Similar results for boundary traction $t_1$ can be found in Fig. 13. By combining the displacements, Fig. 14 illustrates the mechanical deformation of the structure, where the red-line in the background represents the un-deformed shape.

**Table 2.** Comparison of the present and the traditional PINNs methods.

| Reference point (0.9, -0.4) | Exact ($\times 10^{-4}$) | Present BINNs (180 points) | Relative error | Traditional PINNs (2209 points) | Relative error |
|---|---|---|---|---|---|
| $u_1\,(\mathrm{mm})$ | 1.1914 | 1.1914 | $9.6040\times 10^{-6}$ | 1.2196 | $2.3764\times 10^{-2}$ |
| $u_2\,(\mathrm{mm})$ | 1.4264 | 1.4265 | $6.4356\times 10^{-5}$ | 1.4922 | $4.6106\times 10^{-2}$ |
| $\phi\,(\tilde{\mathrm{V}})$ | -7.9897 | -7.9897 | $1.3887\times 10^{-7}$ | -7.9901 | $4.4375\times 10^{-5}$ |
| $\sigma_{11}\,(\tilde{\mathrm{N}}\cdot\mathrm{mm}^{-2})$ | -0.1300 | -0.1300 | $5.8286\times 10^{-5}$ | -0.1293 | $5.6753\times 10^{-3}$ |
| $D_2\,(\tilde{\mathrm{N}}\cdot\tilde{\mathrm{V}}^{-1}\cdot\mathrm{mm}^{-1})$ | -3.0014 | -3.0014 | $9.6159\times 10^{-7}$ | -3.0012 | $5.3261\times 10^{-5}$ |
| CPU-time (s) | | 56.3 | | 172.5 | |

Table 2 presents the numerical results of the present BINNs at the reference point (0.9, -0.4). For the purpose of comparison, numerical results calculated by using the traditional physics-informed neural networks (PINNs) are also provided. Here, the network of the traditional PINNs is trained by using a finer $47\times 47$ rectangular grid and with 2500 iteration steps. It can be observed from Table 2 that the predicted results calculated by using the present BINNs exhibit excellent agreement with the reference results, with the largest relative error is less than $6.5\times 10^{-5}$. In addition, the performance of the present BINNs is always better than the traditional PINNs, even when employing significantly fewer training data, while maintaining an equivalent number of layers and neurons.

Figs. 15-17 show the contour plots of the calculated displacements $u_1$, $u_2$ and electric potential $\phi$ across the entire computational domain. The contour plots of the associated relative errors are also provided, which again clearly demonstrates the accuracy and stability of the present method.

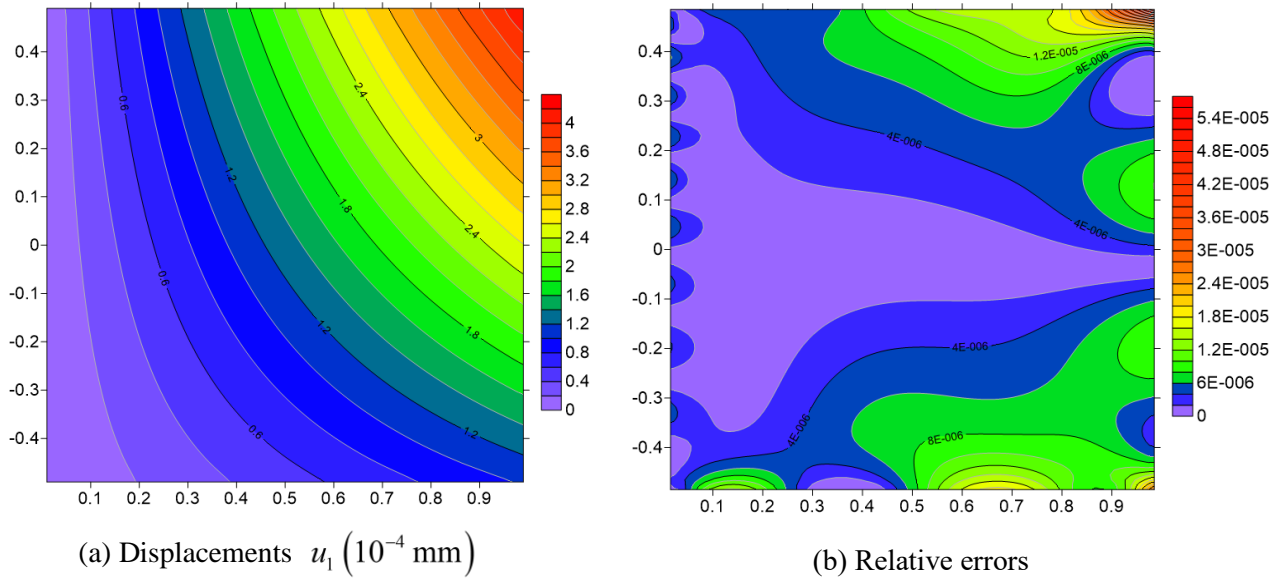

(a) Displacements $u_1$ $(10^{-4}\ \text{mm})$  (b) Relative errors

**Fig. 15.** Contour plots of the displacements $u_1$ (a) and the corresponding relative errors (b).

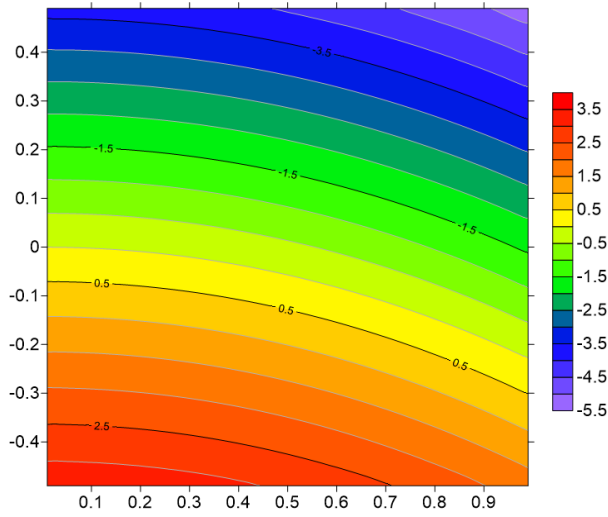
(a) Displacements $u_2 \left(10^{-4} \text{ mm}\right)$

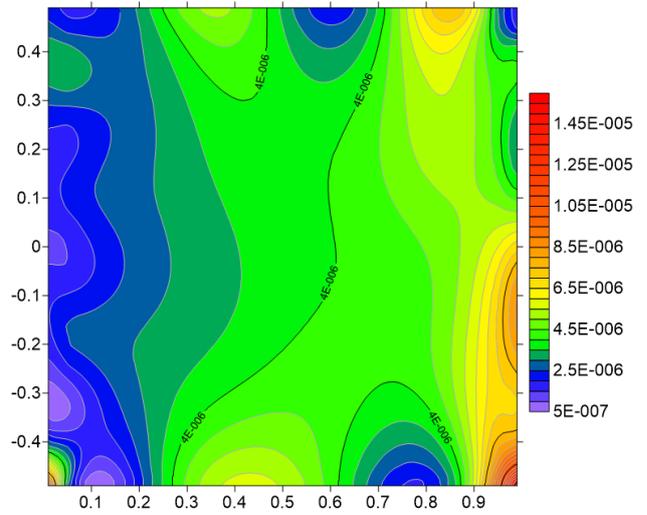
(b) Relative errors

**Fig. 16.** Contour plots of the displacements $u_2$ (a) and the corresponding relative errors (b).

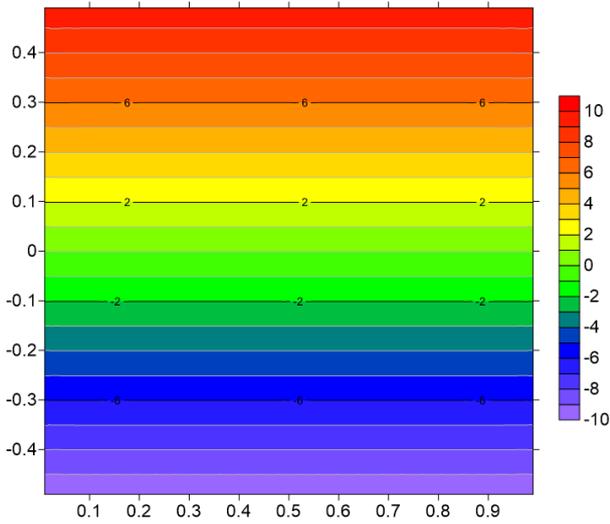
(a) Electric potential $\phi \left(10^{-4} \tilde{\text{V}}\right)$

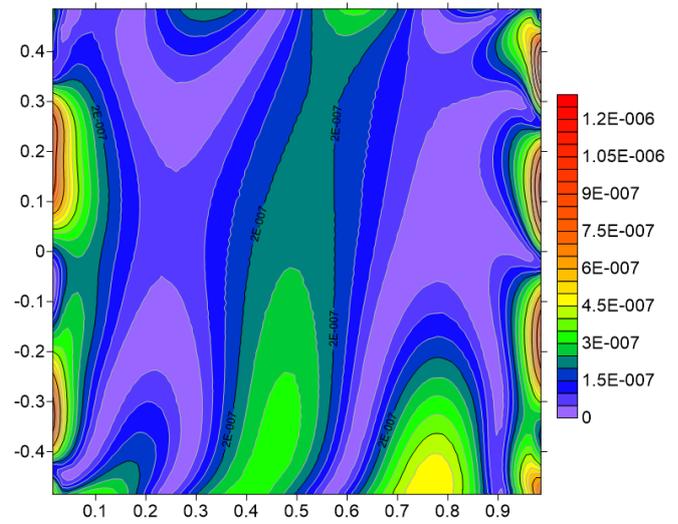
(b) Relative errors

**Fig. 17.** Contour plots of the electric potential $\phi$ (a) and the corresponding relative errors (b).

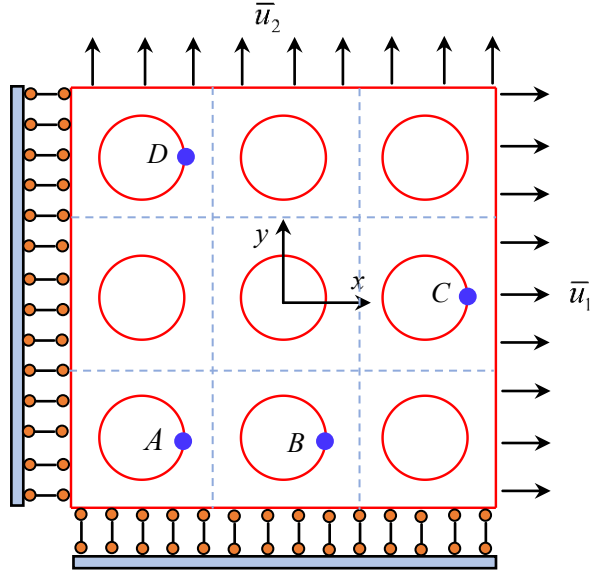

**Fig. 18.** A piezoelectric plane with 9 periodically distributed holes.

### *4.4. A piezoelectric plane with periodically distributed circular holes*

The final example [39, 40], as shown in Fig. 18, considers a $3.0\,\text{m}\times 3.0\,\text{m}$ piezoelectric plane with 9 periodically distributed circular holes of radius $R = 0.3\,\text{m}$. The mechanical and electric constants of the considered piezoelectric plane can be found in Table 3. The boundary conditions of the problem are as follows:

$$u_1 = 0,\quad t_2 = 0,\quad \omega = 0,\quad (left\ boundary), \tag{37}$$

$$u_2 = 0,\quad t_1 = 0,\quad \phi = 0,\quad (lower\ boundary), \tag{38}$$

$$u_1 = 0.003,\quad t_2 = 0,\quad \omega = 0,\quad (right\ boundary), \tag{39}$$

$$u_2 = 0.003,\quad t_1 = 0,\quad \omega = 0,\quad (upper\ boundary), \tag{40}$$

and, on the boundaries of all circular holes, the boundary conditions are traction-free and surface charge $\omega = 0$.

**Table 3.** The mechanical and electric constants of the considered piezoelectric plane.

| Elastic compliance | $s_{11} = 1.282,\ s_{13} = -0.828,\ s_{33} = 1.404,\ s_{55} = 3.906\ (\times 10^{-11}\text{m}^2 \cdot \text{N}^{-1})$ |
|---|---|

| | constants | |
|---|---|---|
| Piezoelectric constants | $d_{31} = -1.917,\ d_{33} = 2.552,\ d_{15} = 4.961\ (\times 10^{-10}\ \mathrm{m\cdot V^{-1}})$ | |
| Dielectric constants | $\zeta_{11} = 1.276,\ \zeta_{33} = 1.047\ (\times 10^{-8}\ \mathrm{N\cdot V^{-2}})$ | |

To solve the problem numerically, 20 discontinuous quadratic elements are distributed along each outer edge and 15 elements along each inner circular boundary. Here, we minimize the loss function through the utilization of 3 fully-connected hidden layers with 20 neurons per layer and 2500 iteration steps. For each hidden layer, the activation function employed is the Sinusoid function, denoted as $\sigma(z) = \sin(z)$. We here selected four reference points, labeled as points A, B, C, and D as shown in Fig. 18, and calculated the physical quantities at these points. Table 4 presents a comparison between the results obtained from the current BINNs and the corresponding results calculated using ANSYS simulations (with 28,800 PLANE-13 FEM elements). Again, the current BINNs results exhibit a significant level of agreement with the results obtained using ANSYS. Fig. 19 shows the contour plots of the calculated displacements $u_1$ and electric potential $\phi$ across the entire computational domain.

**Table 4.** Comparison between the results obtained from the current BINNs and the ANSYS.

| Location | Point A | | Point B | | Point C | | Point D | |
|---|---|---|---|---|---|---|---|---|
| | BINNs | ANSYS | BINNs | ANSYS | BINNs | ANSYS | BINNs | ANSYS |
| $u_1\ (\times 10^{-2}\ \mathrm{m})$ | 0.10799 | 0.10791 | 0.20799 | 0.20792 | 0.30800 | 0.30792 | 0.10799 | 0.10792 |
| $u_2\ (\times 10^{-3}\ \mathrm{m})$ | 0.49997 | 0.49989 | 0.49997 | 0.49997 | 1.49992 | 1.49992 | 2.49984 | 2.49983 |
| $\phi\ (\times 10^{6}\ \mathrm{V})$ | 1.25371 | 1.25357 | 1.25370 | 1.25357 | 3.76113 | 3.76071 | 6.26854 | 6.26787 |

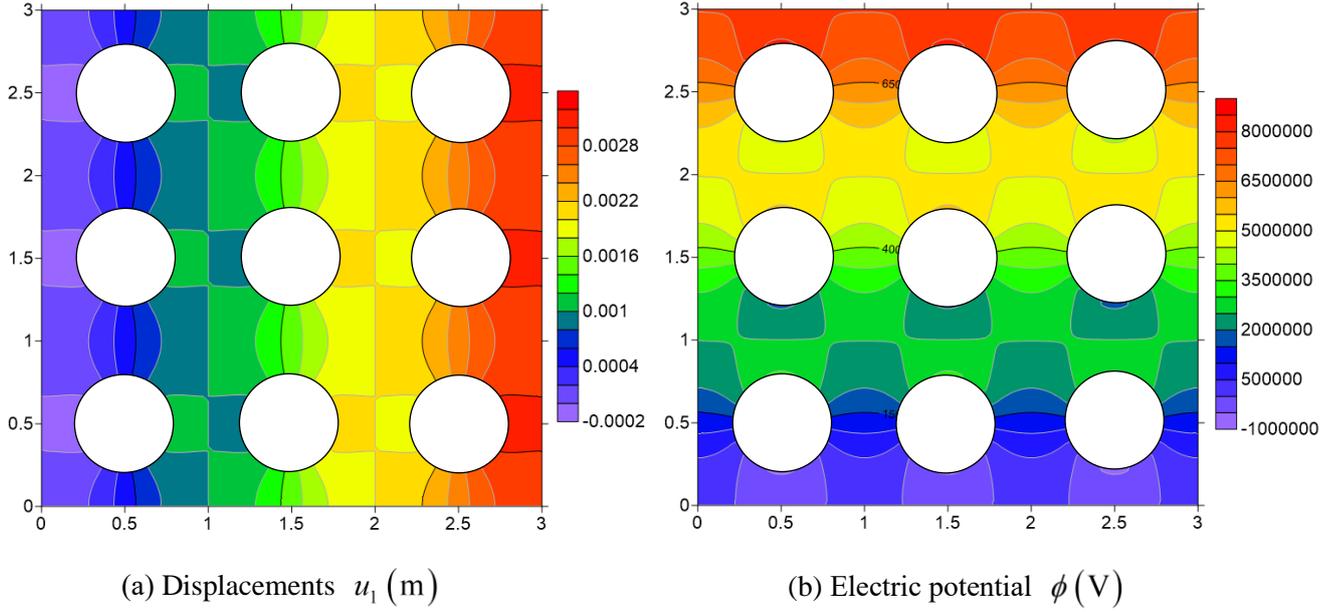

(a) Displacements $u_1$ (m)　　　(b) Electric potential $\phi$ (V)

**Fig. 19.** Contour plots of the displacement $u_1$ (a) and the electric potential $\phi$ (b).

## 5. Conclusions

In this paper, we present a new simulation framework, named as the boundary integrated neural networks (BINNs), for the numerical solution of certain types of boundary value problems. Unlike traditional neural network-based methods that directly solve the original PDEs, the current BINNs incorporates the well-established boundary integral equations (BIEs) of the PDEs into the artificial neural networks. This integration bridges the gap between analytical/semi-analytical methods and data-driven approaches, resulting in more accurate and robust solutions for a wide range of PDE problems. By embedding BIEs into the learning procedure, the current BINNs only need to discretize the boundary of the solution domain and can lead to a faster and more stable learning process than many other neural network-based solvers. The accuracy and efficiency of the present method are demonstrated through several representative benchmark examples in solid mechanics. Finally, although the method exhibits competitive performance compared to traditional neural network-based approaches, adequate and rigorous mathematical analysis of the method are still demanding further careful investigations. The current work establishes a solid basis for further interesting research topics, and the corresponding results will be reported in the future.

## Acknowledgements

The work described in this paper was supported by the National Natural Science Foundation of China (Nos. 11872220, 12111530006), and the Natural Science Foundation of Shandong Province of China (Nos. 2019KJI009, ZR2021JQ02).

## Data Availability Statement

The data that support the findings of this study are available from the corresponding author upon reasonable request.